\documentclass[final,authoryear,5p,times,twocolumn]{elsarticle}
\usepackage{graphicx}
\usepackage{epsfig}
\usepackage{amssymb}

\journal{New Astronomy}
\begin{document}
\begin{frontmatter}
\title{The 2016 outburst of the unique symbiotic star MWC 560 (= V694 Mon), its long-term BVRI evolution and a marked 331 days periodicity}
\author[un,um]{U.~Munari} 
\author[an]{S.~Dallaporta} 
\author[an]{F.~Castellani}
\author[an]{L.~Baldinelli}
\author[an]{G.~L.~Righetti}
\author[an]{M.~Graziani}
\author[an]{G.~Cherini}
\author[an]{A.~Maitan}
\author[an]{S.~Moretti}
\author[an]{S.~Tomaselli}
\author[an]{A.~Frigo}

\address[un]{corresponding author: Tel.: +39-0424-600033, Fax.:+39-0424-600023, e-mail: ulisse.munari@oapd.inaf.it}
\address[um]{INAF Astronomical Observatory of Padova, via dell'Osservatorio 8, 36012 Asiago (VI), Italy}
\address[an]{ANS Collaboration, c/o Astronomical Observatory, 36012 Asiago (VI), Italy}

\begin{abstract}
After 26 years from the major event of 1990, in early 2016 the puzzling
symbiotic binary MWC 560 has gone into a new and even brighter outburst.  We
present our tight $B$$V$$R_{\rm C}$$I_{\rm C}$ photometric monitoring of MWC
560 (451 independent runs distributed over 357 different nights), covering
the 2005-2016 interval, and the current outburst in particoular.  A
stricking feature of the 2016 outburst has been the suppression of the short
term chaotic variability during the rise toward maximum brightness, and its
dominance afterward with an amplitude in excess of 0.5 mag.  Similar to the
1990 event when the object remained around maximum brightness for $\sim$6
months, at the time Solar conjunction prevented further observations of the
current outburst, MWC 560 was still around maximum, three months past
reaching it.  We place our observations into a long term contex by combining
with literature data to provide a complete 1928-2016 lightcurve.  Some
strong periodicities are found to modulate the optical photometry of MWC
560.  A period of 1860 days regulate the occourence of bright phases at
$B$$V$$R_{\rm C}$ bands (with exactly 5.0 cycles separating the 1990 and
2016 outbursts), while the peak brightness attained during bright phases
seems to vary with a $\sim$9570 days cycle.  A clean 331 day periodicity
modulate the $I_{\rm C}$ lightcurve, where the emission from the M giant
dominates, with a lightcurve strongly reminiscent of an ellipsoidal
distortion plus irradiation from the hot companion.  Pros and cons of 1860
and 331 days as the system orbital period are reviewed, waiting for a
spectroscopic radial velocity orbit of the M giant to settle the question
(provided the orbit is not oriented face-on).
\end{abstract} 
\begin{keyword} novae, cataclysmic variables -- symbiotic binaries
\end{keyword}

\end{frontmatter}

\section{Introduction}
\label{}

MWC 560 was first noted in the Mt. Wilson Catalog of emission line objects
by Merrill \& Burwell (1943) as a Be-type star with bright Balmer emission
lines flaked, on the violet side, by wide and deep absorption lines.  The
presence of a cool giant, betrayed by strong TiO absorption bands visible in
the red, was reported by Sanduleak \& Stephenson (1973), who classify the
giant as M4 and confirmed the presence next to the emission lines of deep,
blue-shifted absorptions.  A short abstract by Bond et al.  (1984) informed
that in early 80ies they measured terminal velocities up to $-$3000
km~sec$^{-1}$ in the Balmer absorption components, the absoption profiles
were very complex and variable on timescales of one day, and flickering with
an amplitude of 0.2 mag on a timescale of a few minutes dominated high-speed
photometry.  Interestingly, Bond et al.  mentioned that near H$\alpha$ the
spectrum was dominated by the M giant, and no absorption componet was seen. 
Compared to post-1990 spectra in which the H$\alpha$ absorption is
outstanding and the visibility of the M giant spectrum is confined to
$\lambda$$\geq$6500/7000 \AA, this indicates that, at the time of the
observations by Bond et al.  (1984), the luminosity of the hot component was
significantly lower than typical for the last 25 years.

In keeping up with the very slow pace at which MWC 560 was attracting
interest, even the promising report by Bond et al.  (1984) did not much to
improve upon the anonymity of MWC 560, until all of a sudden, in early 1990,
MWC 560 took the scene for a few months, with a flurry of IAU Circulars and
near-daily reports conveying increasing excitement.  All started when Tomov
(1990) reported on the huge complexity he had observed in the Balmer absorption
profiles on his January 1990 high resolution spectra,
suggesting ``discrete jet-like ejections with a relatively high degree of
collimation and with the direction of the ejection near to the line of
sight".  Feast \& Marang (1990) soon announced that optical photometry
clearly indicated that the object was in outburst, immediately followed by
Buckley et al.  (1990) who reported terminal velocities up to $-$5000
km~sec$^{-1}$ for the Balmer absorptions, upward revised to $-$6500
km~sec$^{-1}$ by Szkody \& Mateo (1990) a few days later.  By the time Maran
\& Michalitsianos (1990) re-observed MWC 560 with the IUE satellite at the
end of April 1990, the paroxysmal phase was ending.

    \begin{table*}[!Ht]
       \caption{Our 2005-2016 $B$$V$$R_{\rm C}$$I_{\rm C}$ photometric
       observations of MWC 560.  The full table is available electronically
       via CDS, a small portion is shown here for guidance on its form and
       content.}
        \centering
        \includegraphics[width=135mm]{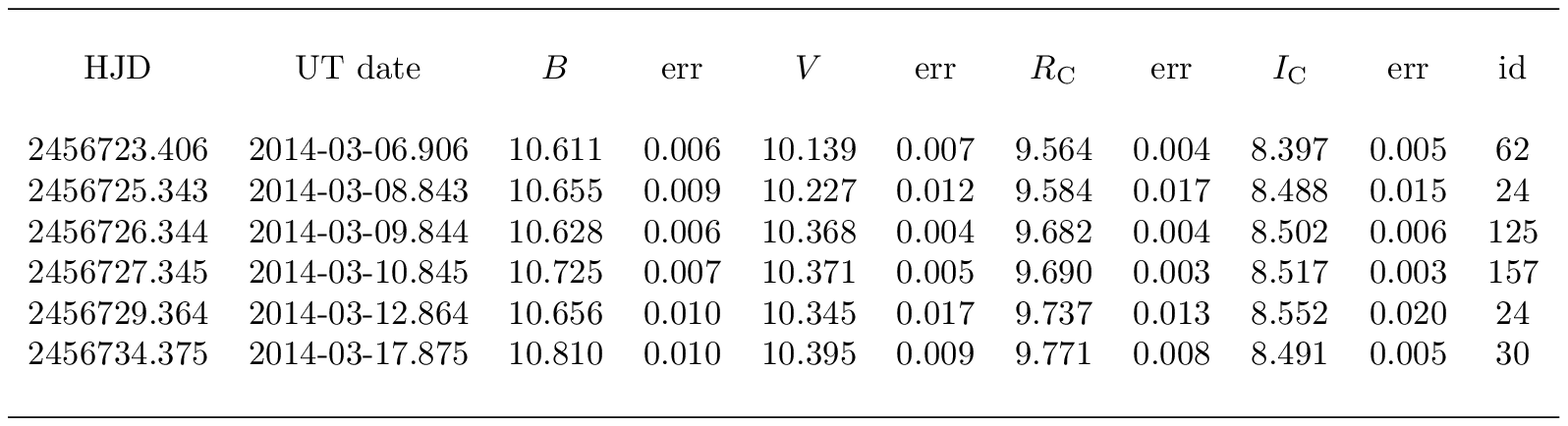}
        \label{tab1}
    \end{table*}
  \begin{figure*}[!Ht]
     \centering
     \includegraphics[angle=270,width=16.2cm]{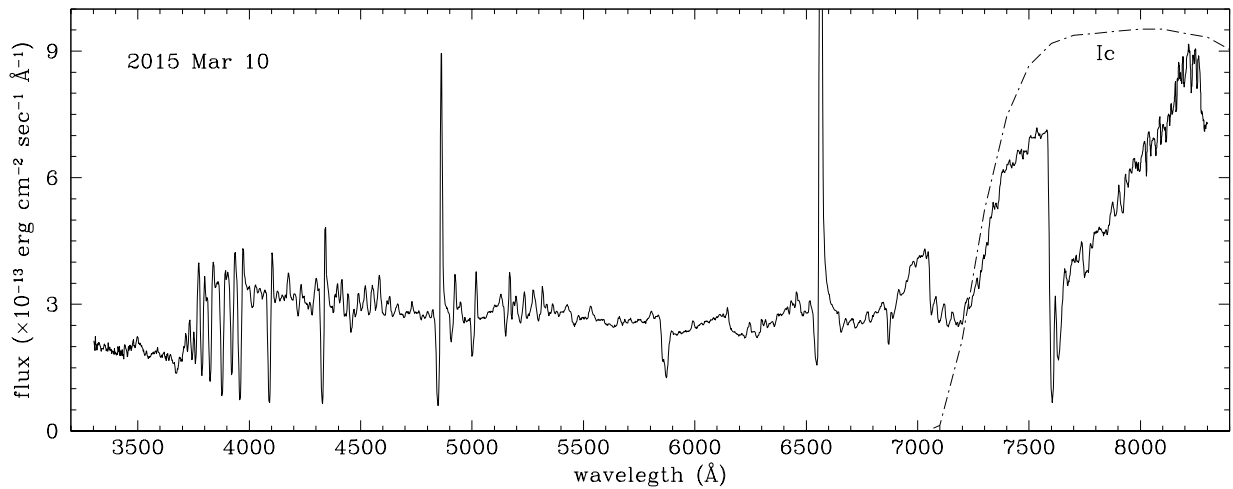}
     \caption{Typical optical spectrum for MWC 560. The dotted-dashed line
     marks the transmission profile of Landolt $I_{\rm C}$ photometric band.}
     \label{fig1}
  \end{figure*}

The nature of MWC 560 as outlined by the 1990 outburst was reviewed by Tomov
et al.  (1990), while the preceeding photometric history tracing back to
1928 was recostructed, from archival photographic plates, by Luthardt
(1991).  A few observations reported by Doroshenko, Goranskij, \& Efimov
(1993) extend the time coverage back to $\sim$1900.  Tomov et al.  (1992)
and Michalitsianos et al.  (1993) modelled MWC 560 with a non-variable M4
giant and an accreting - and probably magnetic - white dwarf (WD),
surrounded by an (outer) accretion disk, and subject to a steady optically
thick wind outflow and a complex pattern of mass ejection into discrete
blobs.  Stute \& Sahai (2009) deduced however a non-magnetic WD from their
X-ray observations.  A fit with a variable collimated outflow that
originates at the surface of the accretion disk and that is accellerated with far
greater efficiency than in normal stellar atmospheres was considered by
Shore, Aufdenberg, \& Michalitsianos (1994).  The collimated jet outflow was
also investigated by Schmid et al.  (2001).  A strong flickering activity
has been present at all epochs in the photometry of MWC 560, with an
amplitude inversely correlated with then system brightness in the $U$ band
(Tomov et al.  1996).  A search for a spectroscopic counterpart of the
photometric flickering was carried out by Tomov et al.  (1995) on high
resolution and high S/N spectra taken during 1993-1994 when the object was
in a quiescent state.  In spite of the very large amplitude of the
photometric flickering recorded on simultaneous $B$$V$ observations
($\sim$0.35 mag in $B$, $\sim$0.20 mag in $V$), no significant change in
intensity and profile (at a level of a few \%) was observed both for the
emission lines and their deep and wide absorption components.

With MWC 560 at quiescence and not much going on with its photometric and
spectroscopic behavior, the interest in the object progressively declined
after the 1990 outburst.  The situation could now reverse following our
recent discovery (Munari et al.  2016) that MWC 560 is going through a new
outburst phase, {\it brighter} than that of 1990.  This has immediately
prompted deep X-ray observations by Lucy et al.  (2016) that found a
dramatic enhancement in the soft ($<$2 keV) X-rays, compared to the
observations by Stute \& Sahai (2009) obtained in 2007 when MWC 560 was in
quiescence.  The report on the optical outburst also prompted VLA
observations that detected for the first time radio emission from MWC 560
(Lucy, Weston, \& Sokoloski 2016), at least an order of magnitude enhanced
over a VLA non-detection on 2014 October 2, during the quiescence preceding
the current outburst.

  \begin{figure*}[!Ht]
     \centering
     \includegraphics[width=17.0cm]{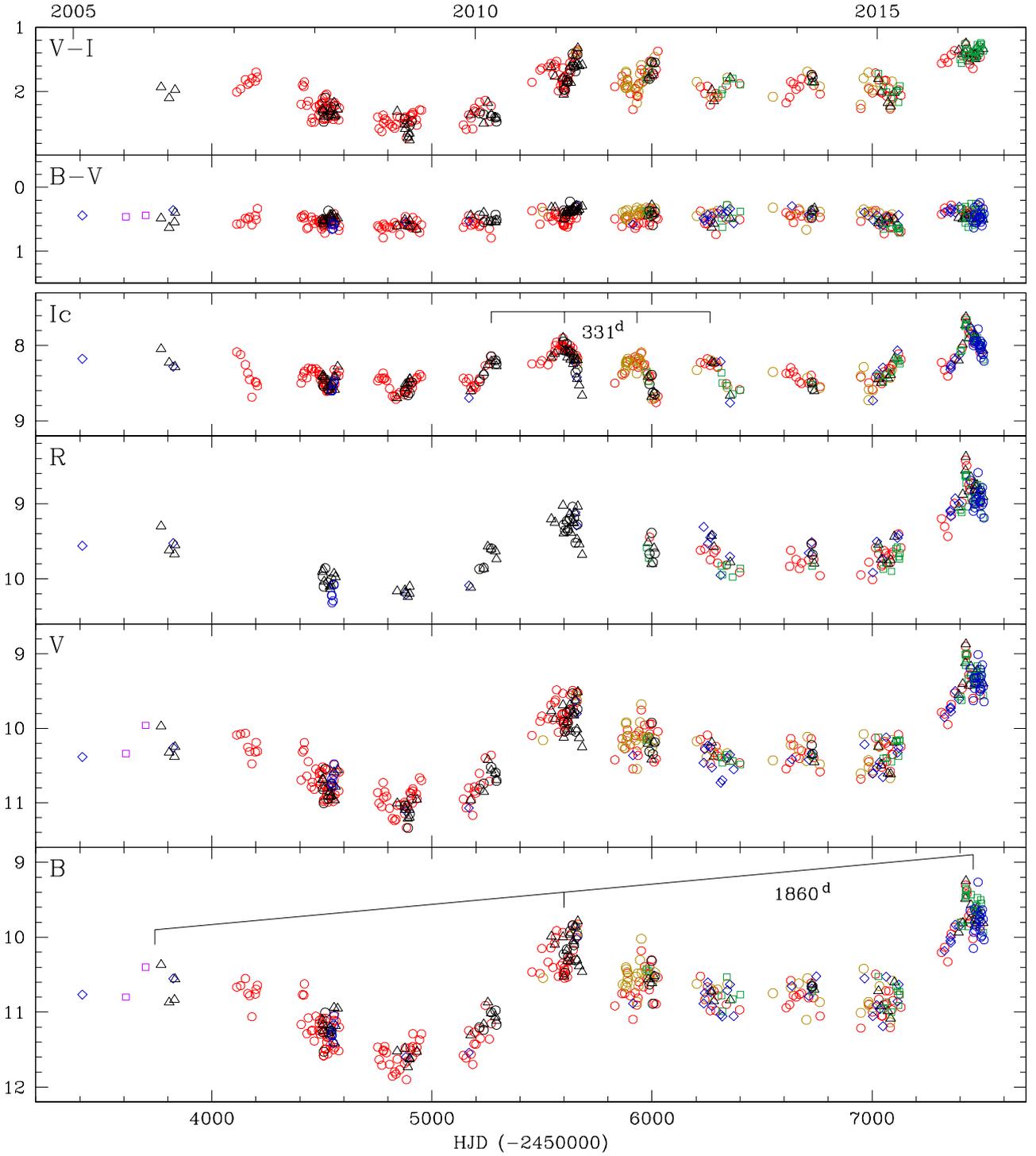}
     \caption{Light and color evolution of MWC 560 from our 2005-2016
     $B$$V$$R_{\rm C}$$I_{\rm C}$ observations. The varied symbols identify 
     different telescopes.}
     \label{fig2}
  \end{figure*}

In this paper we present the results of our 2005-2016 $B$$V$$R_{\rm
C}$$I_{\rm C}$ photometric monitoring of MWC 560, with an emphasis on the
current outburst phase.  This is placed into an historical context by
combining with existing data that provides an optical lightcurve of MWC 560
covering almost a century.  Finally, a search for periodicities is carried
out, especially taking advantage of our unique set of $I_{\rm C}$ data
which is dominated by the emission from the M giant.

\section{Observations}

$B$$V$$R_{\rm C}$$I_{\rm C}$ optical photometry of MWC 560 is regularly
obtained since 2005 with nine of the ANS Collaboration telescopes, all of
them located in Italy.  A total of 431 $B$$V$$R_{\rm C}$$I_{\rm C}$
independent runs are presented here, obtained during 357 different nights
distributed between Feb 9, 2005 and Apr 29, 2016.  The operation of ANS
Collaboration telescopes is described in detail by Munari et al.  (2012) and
Munari \& Moretti (2012).  The same local photometric sequence, calibrated
by Henden \& Munari (2001) against Landolt equatorial standards, was used at
all telescopes on all observing epochs, ensuing a high consistency of the
data.  The $B$$V$$R_{\rm C}$$I_{\rm C}$ photometry of MWC 560 is given in
Table~1, where the quoted uncertainty is the total error budget, which
quadratically combines the measurement error on the variable with the error
associated to the transformation from the local to the standard photometric
system (as defined by the photometric comparison sequence).  All
measurements were carried out with aperture photometry, the long focal
length of the telescopes and the absence of nearby contaminating stars not
requiring to revert to PSF-fitting.

  \begin{figure*}[!Ht]
     \centering
     \includegraphics[angle=270,width=17cm]{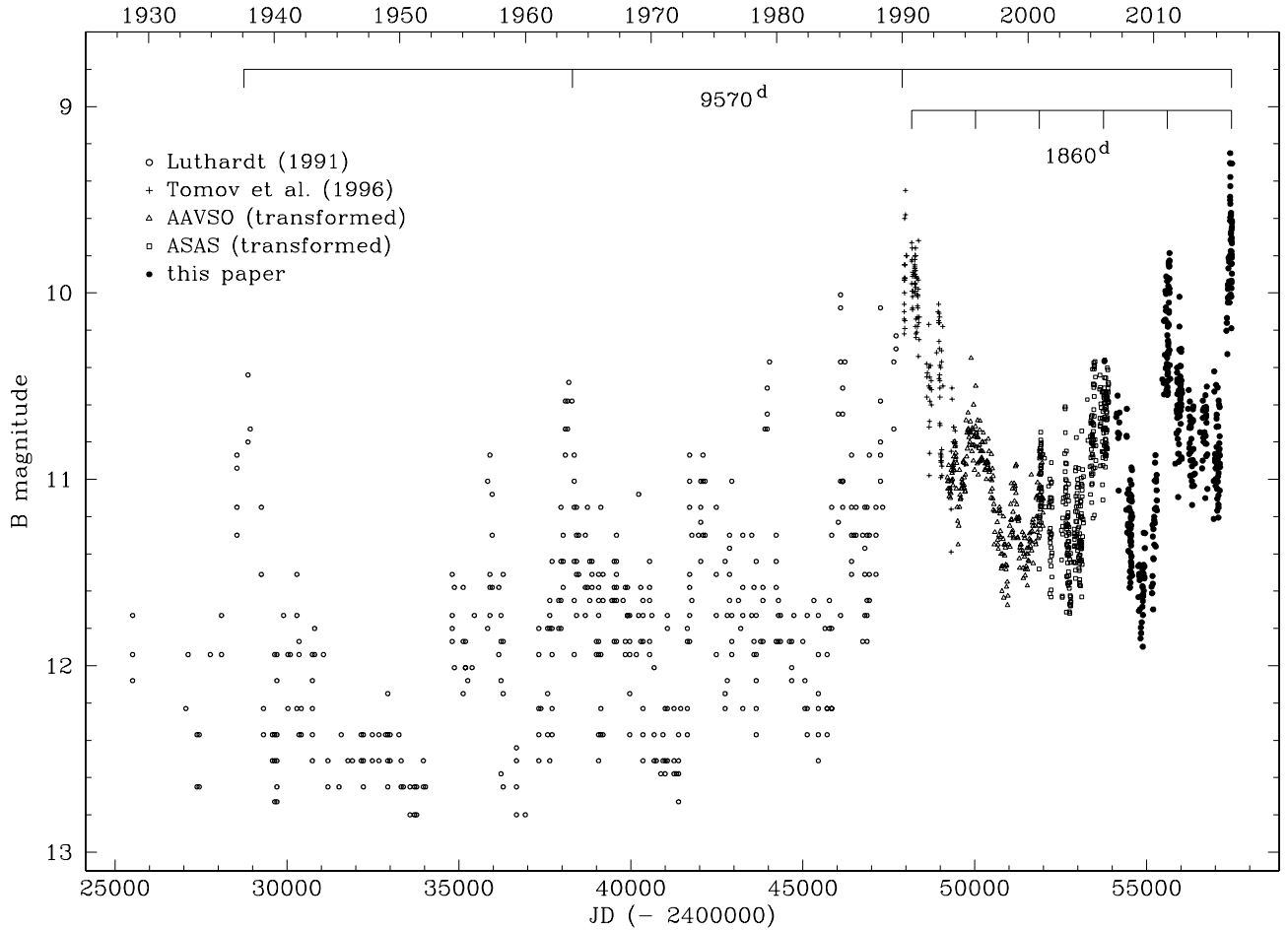}
     \caption{Long term photometric evolution of MWC 560 in the $B$ band.} 
     \label{fig_hist}
  \end{figure*}

Low and high resolution spectra of MWC 560 are routinely obtained with the
Asiago 1.22m + B\&C and 1.82m + REOSC Echelle, and with Varese 0.61m +
Astrolight mk.III multi-mode spectrograph.  One of the low resolution
spectra is shown in Figure~1 as representative of the typical appearance of
MWC 560, broadly similar both in quiescence and outbursts, with the
continuum from the M giant becoming rapidly dominant for
$\lambda$$>$7000~\AA\ and over the Landolt's $I_{\rm C}$ band. The results
of the spectroscopic campaign on MWC 560 will be discussed elsewhere.

\section{Long term photometric evolution}

The $B$$V$$R_{\rm C}$$I_{\rm C}$ lightcurve of MWC 560 covering the last
eleven years is presented in Figure~2.  This time interval corresponds to
slightly more than two full cycles of the 1900$\div$2000 days periodicity
frequently associated to MWC 560 (see sect.  5 below).  The presence
of such a periodicity ($\simeq$5.2 years) is evident in the $B$-band panel
of Figure~2, where the maxima of 2006, 2011, and 2016 clearly modulate the
lightcurve.  Contrary to what found by others (eg.  Tomov et al.  1996,
their Figure~1), our data in Figure~2 show that the $B$$-$$V$ color of MWC
560 remains essentially stable in spite of the large changes recorded in
$B$ band.  A variation at the level of $\sim$1 mag is instead observed in the
$V$$-$$I_{\rm C}$ color, with MWC 560 being redder when fainter at $B$.  The
lightcurve of MWC 560 in the $I_{\rm C}$ band is completely different from
those at shorter wavelengths, in particular: ($a$) the large scatter that
dominates the $B$ lightcurve, which is caused by the accretion flickering,
is null at $I_{\rm C}$, ($b$) the large amplitude maxima of 2006, 2011, and
2016 that dominates the $B$-band lightcurve are barely recognizable at
$I_{\rm C}$, and ($c$) a clear, large amplitude ($\sim$0.35 mag) and
periodic modulation governs the $I_{\rm C}$ lightcurve.

The above is consistent with a hot component that completely dominates the
emission at $B$ and $V$, and contributes the majority of the flux at $R_{\rm
C}$, while the M4 giant accounts for nearly all the brightness recorded
in $I_{\rm C}$.  The hot component, presumably a massive accretion disc
around the WD companion, is the one producing the continuum mimicking an
A-type star which dominates shortward of 6000 \AA\ in the spectrum of
Figure~1, while the TiO bands that dominate longward of 7000 \AA\ come from the
M4 giant.  The M4 giant is intrinsically variable, and entirely responsible
for the periodic changes seen in $I_{\rm C}$.  They could either be the
result of an ellipsoidal distortion (in which case the M4 giant would fill
the corresponding Roche lobe) or be caused by a radial pulsation (in which
case the M4 giant would {\it not} fill its Roche lobe; see sect.  5).  The
flickering that affects the emission from the hot component causes a large
dispersion of the observations at $B$$V$$R_{\rm C}$ bands, much larger than
the observational errors.  The dispersion is significantly lower in the
$B$$-$$V$ color (see Figure~2), because the flickering affects in
phase both bands.

To put things into context, in Figure~3 we combine our 2005-2016 $B$-band
photometry with 1928-1990 $m_{\rm pg}$ photographic photometry from Luthardt
(1991, with its zero point set according to Doroshenko, Goranskij, \& Efimov
1993) and the 1990-1995 $B$-band photometry by Tomov et al.  (1996).  To
fill the remaining 1995-2005 gap, we have used $V$-band ASAS CCD photometry
and AAVSO visual estimates, both transformed to $B$ band, a safe
approximation considering the constant $B$$-$$V$ color displayed by MWC 560
throughout active and quiescence states.  The AAVSO visual estimates were
first averaged into 10-days bins to filter out the intrinsic noise, and a
constant $-$0.1 mag was added to correct their zero point to that of
$V$-band CCD observations.  Finally, a $B$$-$$V$=+0.456 color (corresponding
to the median value of our observations) was added to both ASAS and AAVSO
data to transform them into $B$ values.

The $B$-band historical lightcurve in Figure~3 shows that the 1990
outburst marked a transition in the mean brightness of MWC 560, its
quiescence value being $\sim$12 before and $\sim$11 mag after. 
Estimates on a few older photographic plates by Doroshenko, Goranskij, \&
Efimov (1993), suggest that the brightness of MWC 560 was $\sim$13 before
the 1928-1990 period covered by Luthardt (1991), which would indicate a
marked secular trend toward brighter magnitudes superimposed to an always
present large variability over a wide range of timescales.

\section{The 2016 outburst}

The record brightness attained by MWC 560 in early 2016 is obvious in the
century-long lightcurve of Figure~3.  A zoomed view of the outburst light-
and color-curves is presented in Figure~4.

The outburst peaked at $B$=9.25 on 2016 Feb 7, and a second maximum was
reached on April 3 at $B$=9.21.  The complete flatness of $B$$-$$V$ and
$V$$-$$R_{\rm C}$ color-curves in Figure~4 indicates that the spectral
energy distribution of the outbursting component did not change, only the
overall intensity varied.  At peak brightness, the outbursting component was
so bright to profusely leak into the $I_{\rm C}$ band and dominate over the
emission from the M4 giant.

A striking feature of the outburst is well visible in Figure~4: on the rise
toward the main peak of Feb 7, the $B$-band lightcurve is much smoother
compared to its post-peak portion which is affected by a $\sim$0.5 mag
scatter.  It is like if the always present flickering was momentarily
suppressed during the rise to maximum, but soon after reaching it,
the flickering re-appeared with an even larger virulence than in quiescence.

Even if MWC 560 has not displayed a flat maximum, nonetheless it has
remained around maximum for the three months covered by our observations
before they were interrupted by the Solar conjunction (cf Figures 2 and 4). 
This is reminiscent of the behavior displayed at the time of the 1990
outburst, when the star lingered around maximum for six months before
entering a sharp decline phase (cf.  Tomov et al.  1996, their Figure~2).

\section{Periodicities in the photometric behavior}

\subsection{1860 days}

  \begin{figure}[!Ht]
     \centering
     \includegraphics[width=7.8cm]{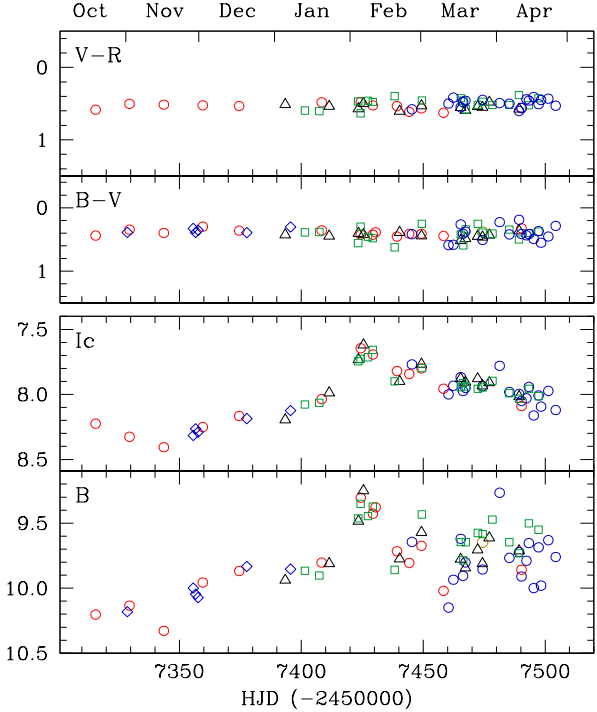}
     \caption{Zoomed view of the portion of Figure~2 covering the
     phase around maximum brightness in early 2016.}
     \label{fig3}
  \end{figure}
 
Some authors have already reported on the presence in the optical lightcurve
of MWC 560 of a period between 1900 and 2000 days, apparently
regulating the occurrence of brightness peaks.  The latter are ascribed to
an increase in mass transfer during passages at periastron.  It should be
noted that a pre-requisite of this scenario is the presence of a highly
eccentric orbit, when most of the symbiotic stars with a spectroscopic orbit
show negligible eccentricities (eg.  Fekel et al.  2000a,b, 2001), and for
only a small fraction of them the eccentricity is significant (eg.  Fekel et
al.  2015).  Going into details, Tomov et al.  (1992) and Frckowiak,
Miko{\l}ajewski, \& Tomov (2003) reported for MWC 560 a period of $\sim$2000
days, Doroshenko et al.  (1993) found 1930 days, while Leibowitz \&
Formiggini (2015) preferred 1943 days.  All these investigations include, as
the main body of data, the brightness estimates that Luthardt (1991) made on
historical photographic plates exposed at Sonneberg Observatory during
1928-1990 (no information is provided on the wavelength sensitivity of the
plates and/or the presence of photometric filters).  Gromadzki et al. 
(2007) noted a period of 1931 days from AAVSO visual estimates covering
1990-2001 and ASAS data distributed over 2000-2005, an interval of time
covering three cycles of the proposed periodicity.

Looking at the 1990-2016 lightcurve of MWC 560 in Figure~3, the time
interval between brightness peaks seems significantly shorter than proposed
in the studies just mentioned.  A period of 1860 days accounts for exactly
5.00 cycles between the brightness peaks in 1990 and 2016, and also nicely
fits the intermediate and less luminous maxima, as shown in Figure~3.
The ephemeris for the 1860 days period is:
\begin{equation}
{\rm max} = 2457460 + 1860 \times E
\end{equation}
Is the P=1860 days period due to orbital motion ? The interpretation in
terms of periastron passages on a highly eccentric orbit is tempting, but
there are some obstacles to accept it.

Doroshenko et al.  (1993) pointed out how MWC 560 has sometimes remained at
minimum when instead a maximum would have been expected, as it was for the
2600 day long interval between 1943 and 1950, the best and densely mapped
period in the historical lightcurve, when MWC 560 remained at a flat minimum
with no trace of cyclic variations, missing two of the expected maxima.  It
is also worth comparing the 1860 day period with the known orbital periods
of symbiotic stars.  In doing this we limit ourselves to the classical
symbiotic stars containing a normal, non-Mira cool giant, as it is the case
for MWC 560.  A survey of the available literature returns validated orbital
periods for $\sim$50 such symbiotic stars.  Their median value is 670 days,
with a 25th percentile of 215 days.  The longest orbital period is 1619 days
for Y CrA.  Thus an orbital period of 1860 days would be unusually long for
a classic symbiotic binary with a normal, non-Mira giant.

Since the first spectroscopic observations by Merrill \& Burwell (1943), the
optical spectrum of MWC 560 has been invariably described similar to that
shown in Figure~1.  This has the characteristics of a very bright (and
therefore large and massive) accretion disk dominating over the emission of
the M4 giant at $\lambda$$<$6000 \AA.  Such a disk must be continuously fed
by massive mass transfer from the M4 giant, as indicated by the reckless and
large amplitude flickering that {\em always} dominates the optical
photometry of MWC 560.  This however requires the M4 giant to fill its Roche
lobe along the whole orbit around the WD, and not just at periastron
passages every $\sim$5 years.

As a final remark on an orbital interpretation for the 1860 days period,
this appears to be somewhat too long for an M4 giant to fill its Roche lobe
in a circular orbit.  Well known examples of symbiotic stars dominated by
ellipsoidal distortions of the Roche-lobe filling giant are: the M3 in T CrB
(orbital period P=227 days), the M4 in QW Sge (P=390 days), the M5 in AX Per
(P=682 days), the M6 in CI Cyg (P=855 days).  This sequence suggests that a
viable orbital period to allow the M4 giant in MWC 560 to fit its Roche lobe
and sustain the observed heavy mass transfer toward the companion should not
exceed 500$-$600 days, with 1860 days well off scale.  An higher-than-usual
luminosity class (and therefore larger radius) for the cool giant in MWC 560
could however complicate this simple picture.

\subsection{9570 days}

The long term behaviour of MWC 560 presented in Figure~3, suggests that a
cycle of $\sim$9570 days (not an integer multiple of 1860) could be
modulating the overall system brightness and that of bright phases, in
particoular during the last 26 years, from the outburst in 1990 to the
present one in 2016.  The two strongest brightness peaks in the historical
data by Luthardt (1991), those in 1937 and 1963, happened in phase with such
a 9570 days cycle.  The corresponding ephemeris is:
\begin{equation}
{\rm max} = 2457460 + 9570 \times E
\end{equation}
The significance of such a period is still preliminary, the known
photometric history of MWC 560 having covered just three such cycles
(cf. Figure~3).

  \begin{figure}[!Ht]
     \centering
     \includegraphics[width=7.8cm]{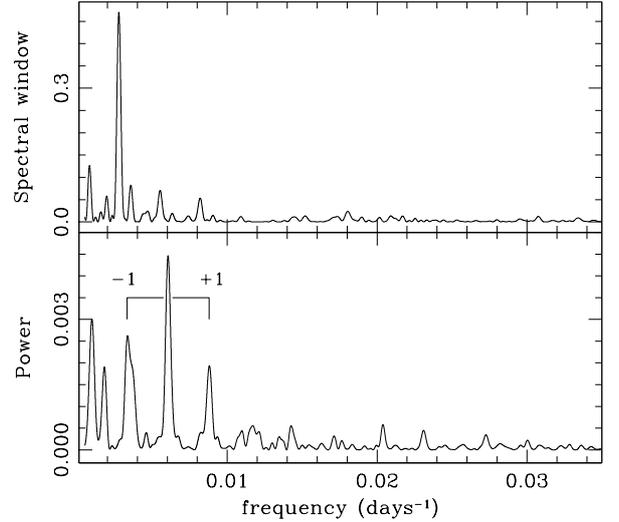}
     \caption{Results of the Fourier analysis of the 2005-2016 $I_{\rm C}$
     data of MWC~560.  The main peak corresponds to a period of 165.4
     days (= 330.8/2, see Eq.~3), flanked by the $+$1 and $-$1 aliases from
     the 365 days sampling frequency (cf. top panel).}
     \label{fig1}
  \end{figure}

  \begin{figure}[!Ht]
     \centering
     \includegraphics[width=7.8cm]{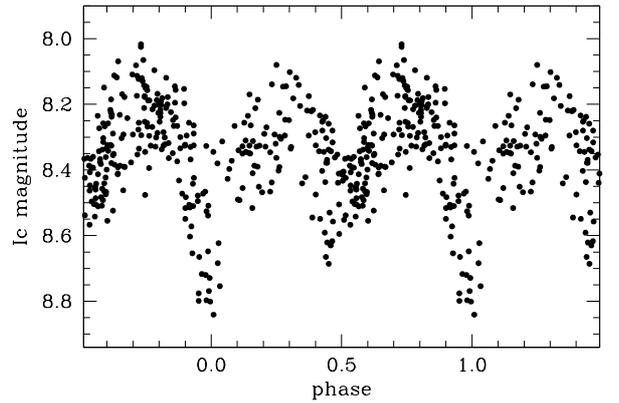}
     \caption{The 2005-2016 $I_{\rm C}$ data from Figure~2 phase plotted
     against Eq.~3 ephemeris (period = 330.8 days).}
     \label{fig1}
  \end{figure}

\subsection{331 days}

The spectrum of MWC 560 in Figure~1 shows the rapid emergence longword of
7000 \AA\ of the TiO bands from the M4 giant.  The $I_{\rm C}$ band centered
at 8200 \AA\ is largely dominated by the emission from the giant, and as
remarked above, eye inspection of the 2005-2016 $I_{\rm C}$ band lightcurve
of MWC 560 in Figure~2 immediately suggests the presence of a periodic
pattern.

We have performed a Fourier search, with the Deeming (1975) code, on our
$I_{\rm C}$ band data in Table~1, excluding the most recent ones affected by
the 2016 outburst.  The resulting periodogram is shown in Figure~5, where
the dominant period is 165.4 days which appears together with the strong
$+$1 and $-$1 year aliases.  The phased lightcurve with twice this period,
or 330.8 days, has a significantly lower dispersion and minima of different
depths, and it is shown in Figure~6.  The ephemeris for primary minima is:
\begin{equation}
{\rm min~(I)} = 2456015 + 330.8 \times E  
\end{equation}
This phased lightcurve has two maxima and two minima per orbital period,
with maxima of similar brightness as in the case of ellipsoidal distortion
of the M giant filling its Roche lobe, and minima on unequal depth as
expected in presence of irradiation - by the hot companion - of the facing
side of the giant.  Should this be indeed the geometry of MWC 560, the above
ephemeris would provide the times of orbital passage at inferior conjunction
of the cool giant.

The amplitude of the modulation in Figure~6 is $\sim$0.35 mag. This value
and the resulting lightcurve is identical to that observed for the symbiotic
binary T CrB (Munari et al.  2016), where an M3 giant orbits the WD
companion every 227.55 days and fills its Roche lobe in a $i$$\sim$68$^\circ$
inclined orbit, as the long term near-IR $J$$H$$K$ lightcurve proves (Yudin
\& Munari 1993).

The presence of a modulation of the far-red and near-IR photometry of MWC 560
has already been noted by Frckowiak, Miko{\l}ajewski, \& Tomov (2003) and
Gromadzki et al. (2007), both reporting periodicities around the 166 and 331
days just discussed for our $I_{\rm C}$ photometry.

Frckowiak, Miko{\l}ajewski, \& Tomov (2003) monitored MWC 560 in
$U$$B$$V$$r$$i$ during 1992-1999, with non-standard $r$ and $i$ bands having
effective wavelengths of 6390 and 7420 \AA\ (for comparison, the
corresponding values for M giants of Landolt's $R_{\rm C}$ and $I_{\rm C}$
bands are 6750 and 8130 \AA).  They found their $i$ band photometry to
behave separately from the other bands, ascribing this to the dominance by
the direct emission of the M giant.  Their Fourier analysis returned a
period of $\sim$161 days.  Unfortunately, Frckowiak, Miko{\l}ajewski, \&
Tomov (2003) did not list or plot their individual observation, but
presented only a phase-{\it averaged} lightcurve.  This does not allow us to
test if - even for their data - a period twice that indicated by the Fourier
analysis would return a better lightcurve, with non-equal minima.

Gromadzki et al.  (2007) reported near-IR $J$$H$$K$$L$ photometry (89
observing dates) of MWC 560 unevenly distributed in time between 1984 and
2004.  Their Fourier analysis returns a period of 339 days, with an
additional peak at 1877 days. They also performed a Fourier analysis of
AAVSO visual estimates + ASAS V-band photometry covering 1990-2006, that
returns a period of 166 days plus its yearly aliases.

\subsection{Other periodicities}

Other possible periodicities, derived from Fourier analysis of mixed MWC 560
data, have been found by single investigations but have not bee reported by
other studies.  They do not seem evident from direct inspection of the
lightcurves.  Doroshenko, Goranskij, \& Efimov (1993) mentioned Fourier
minor peaks at 4570 and 11410 days, Gromadzki et al.  (2007) listed possible
periodicities around 310 and 747 days, and Leibowitz \& Formiggini (2015)
reported about Fourier peaks at $\sim$19000 and 722 days.

\subsection{Which one is the orbital period ?}

Summarizing the results of this section, there are two basic periodicities
observed in MWC 560: 1860 days dominating over $B$$V$$R_{\rm C}$
bands, and 331 days governing the behavior in $I_{\rm C}$.
One of the two is probably the orbital period of the system, but which one 
is not a clean choice.

The difficulties we noted above for the 1860 days being the orbital period
include: ($a$) it is longer than validated known orbital periods for
classical symbiotic stars not harboring a Mira variable, ($b$) the M4 giant
could be too small to fill the corresponding wide Roche lobe along the whole
orbit, a pre-requisite for the uninterrupted, long-term presence of a
massive accretion disk and the always present large amplitude flickering,
($c$) such a bright accretion disk as seen in MWC 560 requires a massive
mass-transfer, as attainable only is a close orbit, in which the donor star
(over)fills its Roche lobe, and ($d$) if the orbital period is 1860 days,
the 331 day should then be ascribed to a pulsation of the giant, but the
shape of the lightcurve is not that typical for a pulsation and the
pulsating periods of non-Mira giants in symbiotic stars are confined between
40 and 200 days (Gromadzki, Miko{\l}ajewska, \& Soszy{\'n}ski 2013).  In
particular, point $b$ point above appears critical, because it would be
incompatible with the supposed high eccentricity orbit that drives the
brightness peaks via enhanced mass transfer during periastron passages,
when the cool giant come in contact with its Roche lobe.

The 331 days modulation of the $I_{\rm C}$ band lightcurve in Figure~6 and
its uneven minima looks attractive.  The shape of the lightcurve suggests a
giant stably filling its Roche lobe and with an irradiated side, a standard
condition for symbiotic stars not harboring a Mira variable.  However, in
order to produce the observed 0.35 mag amplitude for the ellipsoidal
distortion, the orbital inclination of MWC 560 should be relatively high,
$i$$\geq$60$^\circ$, which could contrast with the absence of changes
observed in the radial velocity of emission lines on photographic spectra
obtained at the time of the 1990 outburst (Tomov et al.  1992).  A long-term
check on modern spectra obtained with CCD detectors, of the
constancy/variability in radial velocity of the emission lines in MWC 560
would certainly be worth pursuing.  If the orbital period of MWC 560 is indeed 331
days, it would then be necessary to account for the regularity of the
brightness peaks separated by 1860 days.  For sake of discussion, it could
be speculated that a beating between a non-syncronous axial rotation of the deeply
convective M4 giant and a low orbital eccentricity could perhaps be able to
modulate by a factor of 2$\times$ the mass transfer rate which in turn could
account for the 2$\times$ increase in brightness seen during
active states.

We are forced to conclude that the only {\it robust} way to derive the
orbital period of MWC 560 seems to be a long-term monitoring of the radial
velocity of the M giant, observed as far as possible into the red or near-IR
to contain or null the veiling from the hot companion.  Hoping the orbital
inclination is not anywhere close to $i$=0$^\circ$ and/or there is not much 
radial pulsation to interfere with the signature of orbital motion.

\section{Acknowledgements}
We would like to thank L. Formiggini for communicating in electronic format the
1928-1990 data made available only in graphical form by Luthardt (1991).

\end{document}